\pdfoutput=1

\documentclass[twocolumn,rmp,superscriptaddress,nofootinbib,floatfix]{revtex4}

\usepackage{dcolumn}
\usepackage{graphicx}
\usepackage{rotating}
\usepackage{multirow}
\usepackage{amsmath,amsfonts,amssymb}
\usepackage{algorithmic}
\usepackage{epsfig}
\usepackage{appendix}

\begin{document}

\title{Differences in Impact Factor Across Fields and Over Time}
\author{Benjamin M. Althouse}
\affiliation{Department of Biology, University of Washington, Seattle, WA 98105}
\author{Jevin D. West}
\affiliation{Department of Biology, University of Washington, Seattle, WA 98105}
\author{Theodore Bergstrom}
\affiliation{Department of Economics, University of California Santa Barbara}
\author{Carl T. Bergstrom}
\affiliation{Department of Biology, University of Washington, Seattle, WA 98105}


\begin{abstract}

The bibliometric measure impact factor is a leading indicator of journal influence, and impact factors are routinely used in making decisions ranging from selecting journal subscriptions to allocating research funding to deciding tenure cases. Yet journal impact factors have increased gradually over time, and moreover impact factors vary widely across academic disciplines. Here we quantify inflation over time and differences across fields in impact factor scores and determine the sources of these differences. We find that the average number of citations in reference lists has increased gradually, and this is the predominant factor responsible for the inflation of impact factor scores over time. Field-specific variation in the fraction of citations to literature indexed by Thomson Scientific's Journal Citation Reports is the single greatest contributor to differences among the impact factors of journals in different fields. The growth rate of the scientific literature as a whole, and cross-field differences in net size and growth rate of individual fields, have had very little influence on impact factor inflation or on cross-field differences in impact factor. 

\end{abstract}

%Key words: impact factor, inflation

\maketitle

\section{Introduction}

When Eugene Garfield published his 1972 paper in {\em Science}
describing the role of impact factor in bibliometric studies, he
provided a table of the highest-impact journals in science based
on 1969 data. At that time, only 7 journals had impact factors of
10 or higher, and {\em Science} itself had an impact factor of 3.0
\cite{Garfield72}. Thirty five years later, in  2006,  109
journals have impact factors of 10 or higher, and {\em Science}
registers an impact factor of 30.0 \cite{JCR}. Over the period
from 1994, to 2005, the average impact factor of all  journals
indexed by Journal Citations Reports increased by about 2.6
percent per year.

Average impact factors differ not only over time, but across
fields. For example, in 2006 the highest impact factor in the
field of economics is 4.7, held by the review journal {\em Journal
of Economic Literature}. The top impact factor in molecular and
cell biology is 47.4, held by {\em Annual Reviews of Immunology}.
The average impact factors in these fields differ sixfold: the
average impact factor in economics it is 0.8 whereas the average
in molecular and cell biology is 4.8.

This paper explores the sources of the increase in impact factor
over the past 15 years, and the reasons for impact differences
across fields.  Citation and article counts were obtained from the
CD-ROM version of the Thomson Journal Citation Reports (JCR)
Science and Social Science editions, for the years 1994--2005.

\section{Changes in impact factor over time}
\label{sec:quantifying} A journal's impact factor is a
measure of the number of times that articles published in a census
period cite articles published during an earlier target window.
The impact factor as reported by Thomson Scientific has a one year
census period and uses the two previous years for the target
window. Stated more formally, let $n^i_t $ be the number of times
in year $t$ that the year $t-1$ and $t-2$ volumes of journal $i$
are cited. Let $A^i_t$ be the number of articles that appear in
journal $i$ in year $t$. The impact factor $\text{IF}^i_t$ of
journal $i$ in year $t$ is

\begin{equation}
\text{IF}^i_t=\frac{n^i_t}{A^i_{t-1}+A^i_{t-2}}.
\label{impactfactor}
\end{equation}

\subsection{Impact factors of individual journals}

The JCR database includes 4,300 journals that were indexed continually from
1994 to 2005. For these journals,
 Figure 1a plots 1994 impact factor scores against 2005 scores.
 Points above the diagonal represent journals with impact factor that have risen,
 and points below represent journals with impact factors that have fallen.
 About 80 \% of the journals have increased in impact factor over the eleven years.

\begin{figure*}[htdb]
\centering
\includegraphics[width=\textwidth]{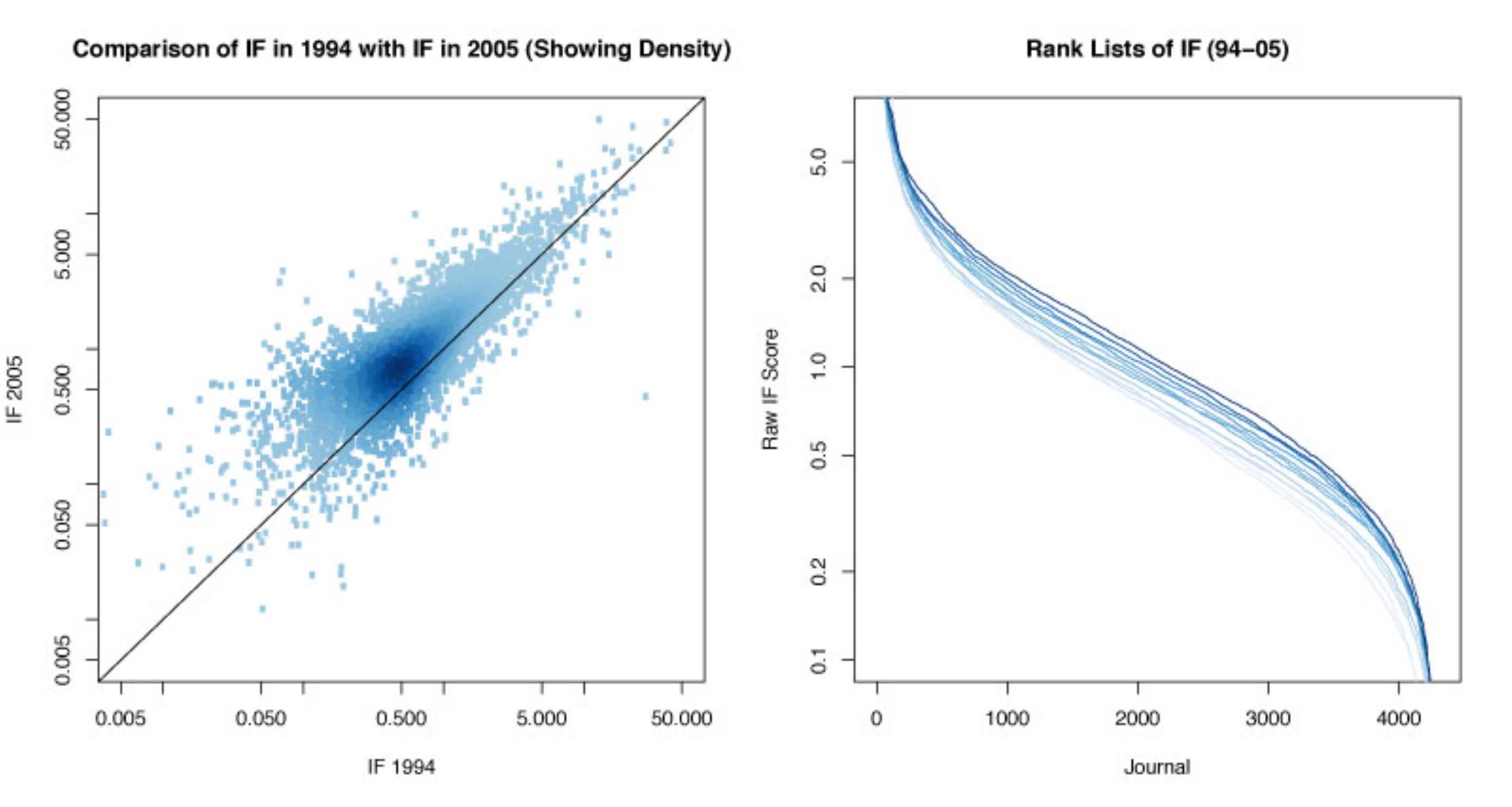}
\caption{{\bf Changes in impact factor from 1994 to 2005.} Panel (a) is a log-log plot of 1994 impact factor against 2005 impact factor for the 4,300 journals that were listed continually from 1994 to 2005 in the JCR.  Shading indicates density, with darker tones representing higher density. Panel (b) plots the rank-order distribution of impact factors from 1994 to 2005. The progression of darkening shade indicates years, with the lightest shade representing 1994 and the darkest 2005.
}
\end{figure*}

Figure 1b shows the rank-order distribution of impact factors  for
years 1994 (lighter blue) through 2005 (darker blue). Impact
factors scores increase annually, predominantly through the
midrange of the distribution. From these figures, it is apparent
that impact factors have increased steadily for most journals,
independently of their initial impact factors.

\subsection{Weighted average impact factor}

To measure average rate of change, it is appropriate to assign
larger weights to journals that publish more articles.  The most
convenient formulation assigns weights proportional to the number of
articles that a journal published during the target years. Let
$A^i_t$ be the number of articles published by journal $i$ in year
$t$ and let $A_t$ be the sum of the articles published over the
set $S_t$ of all journals indexed in year $t$.

We define the weight for journal $i$ in year $t$ as
\begin{equation}w^i_t =\frac{A^i_{t-1}+A^i_{t-2}}{A_{t-1}+A_{t-2}}.\end{equation}
Notice that $\sum_{i\in S_t} w^i_t =1$.  Define the weighted
average impact factor as
\begin{equation}
 \overline{\text{IF}}_t=\sum_{i\in S_t}w^i_t\,\text{IF}^i_t.
\end{equation}

The weighted average impact factor for all journals listed in the
JCR increased by an average rate of  2.6\% percent per year from
1994 to 2005.   For the journals that appeared in the index
throughout the entire period from 1994 through 2005, the average
annual increase was 1.6\%.

\subsection{Decomposing changes in average impact factor}

It might seem appealing to simply
attribute the growth of impact factor to the growth of the
scientific enterprise and in particular to the growth in the number of articles indexed by the JCR.  The raw numbers lend a superficial plausibility to this view.  From 1994 to 2005, the number of articles
in JCR-indexed journals increased by 28\% and the weighted impact
factor increased by 29\%. But with a moment's reflection, we see
that the connection is not immediate.  For any given article, an
increase in the number of related articles is  a source of
additional chances to be cited, but it is also a source of
additional competition for the attention of potential readers and
citations\footnote{This point was observed by Garfield \cite{Garfield06}
who noted that there was no {\em a priori} reason to expect
journals serving large scientific communities  to have higher
impact factors than those serving small ones. }. We will show that a
constant rate of growth of the number of indexed articles will not
result in increasing impact factors unless the number of citations
per article also increases.

We have found a useful way to decompose the average impact factor
in any period into the product of four factors. These are

\begin{enumerate}
\item The number of articles listed in the JCR, as measured by the ratio of number of articles published in the census period to the number of articles published in the target window.
\item The average number of citations in the reference list of each published article.
\item The fraction of all citations from articles written in the census period that cite articles published within the target window of the two prior years.
\item The fraction of cited articles published within the target window that appear in journals indexed by
the JCR.
\end{enumerate}

We construct this decomposition as follows.  Let $c_t$ be the
average number of papers cited by the journals in our dataset (i.e., the JCR-indexed journals)
in year $t$. Let $p_t$ be the fraction of citations in our dataset in year $t$ that go to papers published in
years $t-1$ and $t-2$. Let $v_t$ be the fraction of those citations
appearing in our dataset in year $t$ and referencing items published in years $t-1$ and $t-2$, that go to journals that are listed in the JCR (as opposed to working papers, conference proceedings, books, journals not listed in the JCR, etc.).

Recalling our notation from Section \ref{sec:quantifying},
\begin{equation} \sum_{i\in S_t} n^i_t=A_t\,c_t\,p_t\,v_t,\end{equation}
\noindent and
\begin{eqnarray}
\label{wifderr}
 {\overline{\text{IF}}}_t&=& \sum_{i\in S_t}w^i_t\,\text{IF}^i_t \nonumber \\
 &=&  \sum_{i\in S_t} \frac{A^i_{t-1}+A^i_{t-2}}{A_{t-1}+A_{t-2}} \cdot \frac{n^i_t}{A^i_{t-1}+A^i_{t-2}} \nonumber \\
 &=& \frac{\sum_{i\in S_t}n^i_t}{A_{t-1}+A_{t-2}} \nonumber \\
&=& \frac{A_t\,c_t\,p_t\,v_t}{ A_{t-1}+A_{t-2}}.
\end{eqnarray}

If we  define $\alpha_t=A_t/(A_{t-1}+A_{t-2})$, the weighted
average impact factor at time $t$ can be written as the product
 \begin{equation}\label{product} \overline{\text{IF}}_t=\alpha_t\,c_t\,p_t\,v_t.\end{equation}

The growth rate of a variable is approximated by the change in the
logarithm of that variable. The multiplicative form of equation (\ref{product}) makes it easy to decompose the growth rate of the
average impact factor into the sum of growth rates of the
variables $\alpha$, $c$, $p$, and $v$.
 It follows
from equation (\ref{product}) that
\begin{equation} \rho_t(\overline{\text{IF}})=\rho_t(\alpha)+\rho_t(c)+\rho_t(p)+\rho_t(v),
\label{rhoproduct}
\end{equation}

\noindent where for any variable
$x$, we define $\rho_t(x)=\ln x_t-\ln x_{t-1}$. From the JCR data we are able to determine $\alpha_t$, $c_t$, $p_t$ and $v_t$, and hence $\rho_t(\alpha)$, $\rho_t(c)$,  $\rho_t(p)$ and  $\rho_t(v)$. Our methods for doing so are described in Appendix \ref{Appendix:methods}. The results are reported in Tables \ref{table:acpv} and \ref{table:rhoacpv}.

\begin{table}[htdp]
\centering
\begin{tabular}{c | c | c | c | c | c | c}
Year  ($t$)&\# of articles &$\alpha_t$ & $c_t$ & $p_t$ & $v_t$& $\overline{\text{IF}}$\\
\hline
1994 & 689,876 & 0.544 & 22.121 & 0.176 & 0.835 & 1.764\\
1995  & 709,504 & 0.533 & 22.810 & 0.175 & 0.839 & 1.786\\
1996 & 734,565 & 0.530 & 24.390 & 0.171 & 0.835 & 1.846\\
1997 & 739,890 & 0.517 & 25.040 & 0.167 & 0.833 & 1.796\\
1998  & 753,919 & 0.513 & 27.936 & 0.163 & 0.788 & 1.846\\
1999  & 767,825 & 0.516 & 28.527 & 0.163 & 0.812 & 1.948\\
2000 & 785,583 & 0.518 & 28.913 & 0.162 & 0.820 & 1.988\\
2001  & 788,323 & 0.510 & 29.835 & 0.161 & 0.839 & 2.055\\
2002 & 808,241 & 0.514 & 30.542 & 0.159 & 0.849 & 2.119\\
2003  & 847,705 & 0.535 & 30.666 & 0.157 & 0.857 & 2.206\\
2004  & 885,043 & 0.537 & 31.593 & 0.159 & 0.843 & 2.266

\end{tabular}
\caption{Summary of time behavior of $\alpha_t$, $c_t$, $p_t$ and $v_t$ for the years 1994 to 2004.}
\label{table:acpv}
\end{table}

\begin{table}[htdp]
\centering
\begin{tabular}{c | c | c | c | c | c }
Year ($t$)& $\rho_t(\alpha)$ & $\rho_t(c )$ & $\rho_t(p)$ & $\rho_t(v)$ & $\rho_t(\overline{\text{IF}})$\\
\hline
1995 & -0.019 & 0.031 & -0.004 & 0.005 & 0.012\\
1996 & -0.007 & 0.067 & -0.022 & -0.005 & 0.033\\
1997 & -0.025 & 0.026 & -0.027 & -0.001 & -0.027\\
1998 & -0.007 & 0.109 & -0.019 & -0.056 & 0.027\\
1999 & 0.005 & 0.021 & -0.002 & 0.030 & 0.054\\
2000 & 0.004 & 0.013 & -0.007 & 0.010 & 0.020 \\
2001 & -0.015 & 0.031 & -0.006 & 0.023 & 0.033\\
2002 & 0.008 & 0.023 & -0.013 & 0.012 & 0.031\\
2003 & 0.040 & 0.004 & -0.012 & 0.009 & 0.040\\
2004 & 0.004 & 0.03 & 0.009 & -0.016 & 0.027\\
\hline
Mean & -0.001 & 0.036 & -0.010 & 0.001 & 0.025

\end{tabular}
\caption{Summary of time behavior of $\rho_t(\alpha), \ \rho_t(c ), \ \rho_t(p), \ \rho_t(v)$ and $\rho_t(\overline{\text{IF}})$ for the years 1995 to 2004. The $\rho$ values approximate the fractional annual increase in each component $\alpha$, $c$, $p$, and $v$, and $\overline{\text{IF}}$. The final row shows the average annual increase of each component over the period 1995--2004.}
\label{table:rhoacpv}
\end{table}

Somehow we have to account for an average increase in weighted impact factor of $2.6\%$ per year over the period 1994-2005. Which of the four components is chiefly responsible? Table \ref{table:rhoacpv} lists the $\rho$ values for each component in each year; these $\rho$ values approximate the fractional increase due to each component in each year, and as such, provide the answer.

The increase in the number of articles published over the period 1994--2005 cannot explain the increase in impact factor over the same period. 
The values of $\rho_t(\alpha)$  are small in most years and this component contributes an overall average growth of $-0.001$, i.e., impact factors would decline at an average of roughly $0.1$ percent per year if this were the only factor operating. The basic intuition underlying this result is as follows: first, note that larger fields do not have higher impact factors by shear virtue of their size. While more articles are published in larger fields and thus more citations are given out, those citations are shared among a larger pool of papers.   Second, note that when a field grows at a constant rate, there will be more citation sources published year $t$ than citation targets published in year $t-1$, but this ratio of citation sources to citation targets will remain constant over time, and thus this difference will not inflate impact factors either.

We can show this formally.  Suppose that the number of articles published
grows at a constant rate $\gamma$ and that $c$, $p$, and $v$
remain  constant.  Then $A_t=(1+\gamma)^t$ and hence
\begin{eqnarray}
\alpha_t  &=& \frac{(1-\gamma)^t}{(1-\gamma)^{t-1} + (1 - \gamma)^{t-2}} \nonumber \\
&=&\frac{(1+\gamma)^2}{2+\gamma}. \label{alph}
\end{eqnarray} Since $\alpha_t$ is constant, $\rho_t(\alpha)=0$
for all $t$.

Thus a constant rate of growth, $\gamma$, in the number of articles indexed annually leads to a constant impact factor (no inflation). However, higher rates of growth will yield higher constant impact factors because the derivative of equation (\ref{alph}) with respect to $\gamma$ is positive. By contrast, accelerating growth in the number of articles published (increasing $\gamma$ over time) generates impact factor inflation and decelerating growth generates impact factor deflation. Changes in the other model parameters influence impact factor in more straightforward fashion. Likewise, increasing the average number of outgoing citations per article generates a corresponding increase in impact factor. Increasing the fraction of citations into the measurement window (the fraction of citations to JCR-indexed literature in years $t-1$ and $t-2$) generates a corresponding increase in impact factor. 

We cannot explain the increase in weighted impact factor by means of the change in the fraction of articles citing papers published within the recent two years. The $\rho_t(p)$ values are almost all negative, and in fact this component reduces the impact factor by an average of 1 percent per year over the period 1994--2005. In other words, impact factors would drop considerably if this were the only factor operating. Nor can we explain the increase in impact factor by changes in the fraction of citations to indexed articles. The average $\rho_t(v)$ is only 0.001, i.e., the increase in the fraction of citations that reference indexed articles contributes only about $0.1$ percent per year to the increase in impact factor.  

This leaves as an explanation the change in the number of times the average article cites  a reference source. Table \ref{table:acpv} reveals a  monotone increase in the average number of reference items cited ($c_t$), and in Table 2 we see that this contributes large positive $\rho_t(c)$ values in each year, such that the average increase is approximately 3.6 percent, which adequately explains the 2.6\% increase in weighted impact factor despite the net decline in due to the other components. In short, as citation practices change over time, the average number of citations per article is increasing, and the results is an inflation in impact factor over time.

\bigskip

\bigskip

Our analysis indicates that the single greatest contributor to impact factor inflation over the period 1994--2005 has been an increase in the average number of references per paper. One can imagine a number of potential causes for this increase. These include:
\begin{enumerate}
\item As the size of a field increases, the number of published papers that are relevant to any given manuscript might be expected to increase. Thus we might expect reference lists to grow longer as fields get bigger.
\item Internet search engines, on-line citation databases, and electronic access to the literature itself have considerably reduced the time-cost to authors of finding and obtaining relevant articles. This may have resulted in a concomitant increase in the number of cited items.
\item As researchers become increasingly aware of the value of citations to their own work, referees may demand that authors add numerous citations to their work, and authors may preemptively cite any number of potential editors and referees in their manuscript.
 \end{enumerate}

Preliminary regression analysis provided no evidence that increasing numbers of citable articles lead to increases in the length of reference lists. While it would interesting to seek out data that would allow us to distinguish among the other sources for the change in the average number of references per paper, we do not do so here. 

\subsection{Natural Selection?}

During the period 1994-2005, the JCR added 4,202 new journals that were not previously listed and removed 2,415 journals that were listed in 1994. What effect, if any, did this process of journal substitution have on average impact factors? If the  average impact factors of entering journals exceeded the average impact factor of exiting journals by a sizable margin, this could pull up the entire distribution. We could view this effect as a form of natural selection: the most fit -- those with the highest impact factor scores -- would enter or stay in the data set, while the least fit -- those with the lowest scores -- would drop out of the data set. 

At first glance this seems to be plausible explanation. The journals that enter the JCR over the period 1995--2004 have significantly higher impact factor scores than those that exit over the same period (two sample Kolmogorov-Smirnov test, D = 0.074, $p$ = 5.6e-7). However, even the entering journals had average impact factors well below the average for the full JCR. Because nearly twice as many journals entered as exited, the net effect of flux into and out of the JCR was actually to decrease the average impact factor of the full set of JCR listed journals.  

We see this as follows. For a given year $t$, if we multiply the numbers of articles in years $t-1$ and $t-2$ by the overall weighted impact factor score for that year we can calculate the expected number of citations the set of entering or exiting journals would have to accrue in order to leave the average impact factor of the full set unchanged. The difference between the expected and the actual number of citations brought in by the entering journals can be considered a ``citation cost" of adding new journals (whether positive or negative), and similarly the difference between the actual and the expected number of citations by journals exiting can be considered a ``citation gain" of removing these journals from the data set. We can calculate then, the total effect of the flux of journals in and out of the data set by summing these quantities. For the years 1995--2004, an average cost of 18,200 citations per year was incurred due to turnover in the journals listed. Thus natural selection has not contributed to impact factor inflation.

While the journals that entered the JCR did not on average contribute to impact factor inflation by virtue of entering, they did contribute in the sense that subsequent to entering, their impact factors grew more strongly than did the average for the JCR as a whole. The average annual growth rate for those journals entering in years 1995--2004 is 6\%, more than twice the rate of the overall data set (see also \cite{Wilson07}). Thomson is clearly selecting journals which are rising stars for inclusion in the JCR.

\section{Differences in Impact factor across fields}

Impact factors are well known to vary widely across disciplines \cite{Vinkler88, Seglen97}. Sources of this variation include differences in citation practices \cite{Moed85}, differences in the lag time between publication and subsequent citation  (what we call $p$) \cite{Marton85,Moed85}, and differences in the proportions of citations directed to JCR-indexed literature (what we call $v$) \cite{Hamilton91, Vanclay06}. Here we explore the source of these differences in detail. To delineate disciplinary boundaries, we use the field categories developed by Rosvall and Bergstrom \cite{RosvallBergstrom2008}. These categories use citation patterns to partition the sciences and social sciences into 88 non-overlapping fields.

Table \ref{fieldcompare} lists the 2004 weighted impact factors for the 50 largest fields.  Indeed we see wide variation. For example, the field of Mathematics has a weighted impact factor of $\overline{\text{IF}}=0.56$ whereas Molecular and Cell Biology has a weighted impact factor of 4.76 --- an eight-fold difference. There are several possible sources of this difference, including but not limited to differences in growth rates, differences in the time course of citations, and differences in the fraction of citations that go to non-indexed literature. By extending the model developed in the previous section to partition the weighted impact factor into four separate contributing components, we can quantify the influence of each upon the cross-field differences. 

To begin the analysis we recall Eq. (\ref{rhoproduct}):  \begin{equation*} \rho_t(\overline{\text{IF}})=\rho_t(\alpha)+\rho_t(c)+\rho_t(p)+\rho_t(v). \end{equation*} If journals received citations only from other journals in the same field, the following equation would hold exactly for each field $F$. 
\begin{equation} \rho_t(\overline{\text{IF}}_F)=\rho_t(\alpha_F)+\rho_t(c_F)+\rho_t(p_F)+\rho_t(v_F)
\label{fieldrho}
\end{equation}
In practice, not all citations come from within the same field, so the equation above is only approximate --- though it will be a very good approximation if most cross-disciplinary citations go between fields with similar $\alpha_F$, $c_F$, $p_F$, and $v_F$ values. 

This will let us examine the influence on $\overline{\text{IF}}$ of each component, $\alpha, \ c,\ p, $ and $v$,  in each field $F$ separately. How important is each component? A univariate linear regression of $\rho_t(\alpha)$, $\rho_t(c)$, $\rho_t(p)$,  and $\rho_t(v)$ with $\rho_t(\overline{\text{IF}})$ yields the following coefficients of determination ($r^2$ values, indicating the proportion of total variability explained by each term): 

\begin{eqnarray} \label{rsquaredonevar}
r^2_\alpha = 0.045\nonumber \\
r^2_c =0.172\nonumber \\
r^2_p =0.083\nonumber \\
r^2_v =0.456 
\end{eqnarray}

\noindent These coefficients of determination tell us a number of things. Firstly, the low value of $r^2_\alpha$ indicates that $\alpha_t$, the total number of articles in year $t$ over the total numbers of articles in years $t-1$ and $t-2$, explains very little of the variance across fields weighted impact factor. In contrast, the high value of $r^2_v$ indicates that  the fraction of citations that go into ISI-listed material, $v_F$, explains the greatest fraction of variation of any of the four components. 

If we progress to a multiple regression among pairs of variables, we find:

\begin{eqnarray} \label{rsquaredtwovar}
r^2_{\alpha, c} =0.235 \nonumber \\
r^2_{\alpha, p} =0.118\nonumber \\
r^2_{\alpha, v} =0.457 \nonumber \\
r^2_{c,p} =0.401\nonumber \\
r^2_{c,v} = 0.585 \nonumber \\
r^2_{p,v} = 0.577
\end{eqnarray}

\noindent This further demonstrates the minimal explanatory power of $\alpha$: $r^2_{\alpha,v}$ is approximately equal to $r^2_v$, and similarly for $r^2_{\alpha,c}$ and $r^2_{\alpha,p}$. It also confirms the considerable predictive power of $v$ -- any regression containing $v$ has a relatively high $r^2$, and shows that $c$ and $p$ are also predictively useful in concert with $v$. Multiple regressions with three and four variables yield:

\begin{eqnarray} \label{rsquaredthreevar}
r^2_{\alpha,c,p} = 0.451\nonumber \\
r^2_{\alpha,c,v} = 0.591\nonumber \\
r^2_{\alpha,p,v} = 0.577\nonumber \\
r^2_{c,p,v} =0.854\nonumber \\
r^2_{\alpha,c,p,v} =0.855
\end{eqnarray}

\noindent The $r^2$ with all four variables is 0.855; the model is unable to perfectly predict the weigted impact factor because our assumption that all citations received come from the same field is not strictly true. Noice also that $r^2_{\alpha,c,p,v} \cong r^2_{c,p,v}$, further indicating that $\alpha$ has little, if any, predictive power. 

The method of Hierarchical Partitioning \cite{ChevanSutherland1991} provides a more formal method to estimate the relative contributions or ``importance" of the various independent variables in explaining the total explained variance in a multivariate regression.  The statistic $I$ estimates the contribution of each independent variable. Using the hierarchical partitioning {\tt hier.part} package by Chris Walsh in the statistical analysis program R, we find the following $I$ values for the year 2004 data.

\begin{table}[htdp]
\begin{center}
\begin{tabular}{c|c}
\text{Predictor} &I (\%)\\
\hline
$\alpha$ & 2.858\\
$c$ & 26.624\\
$p$&20.178\\
$v$&50.340
\end{tabular}
\end{center}
\label{ipercents}
\end{table}

\noindent These results indicate that the predictor $v$ (the fraction of citations  to JCR-indexed literature) accounts for 50\% of the explained variance $\overline{\text{IF}}$. The predictor $c$ (number of outgoing citations per article) accounts for an additional 27\%. Those fields which cite heavily within the ISI data set, such as Molecular Biology or Medicine, buoy their own scores. Those fields which do not cite heavily within the ISI data set such as Computer Science or Mathematics have correspondingly lower scores.

\begin{figure*}[htdb]
\centering
\includegraphics[width=3.3 in]{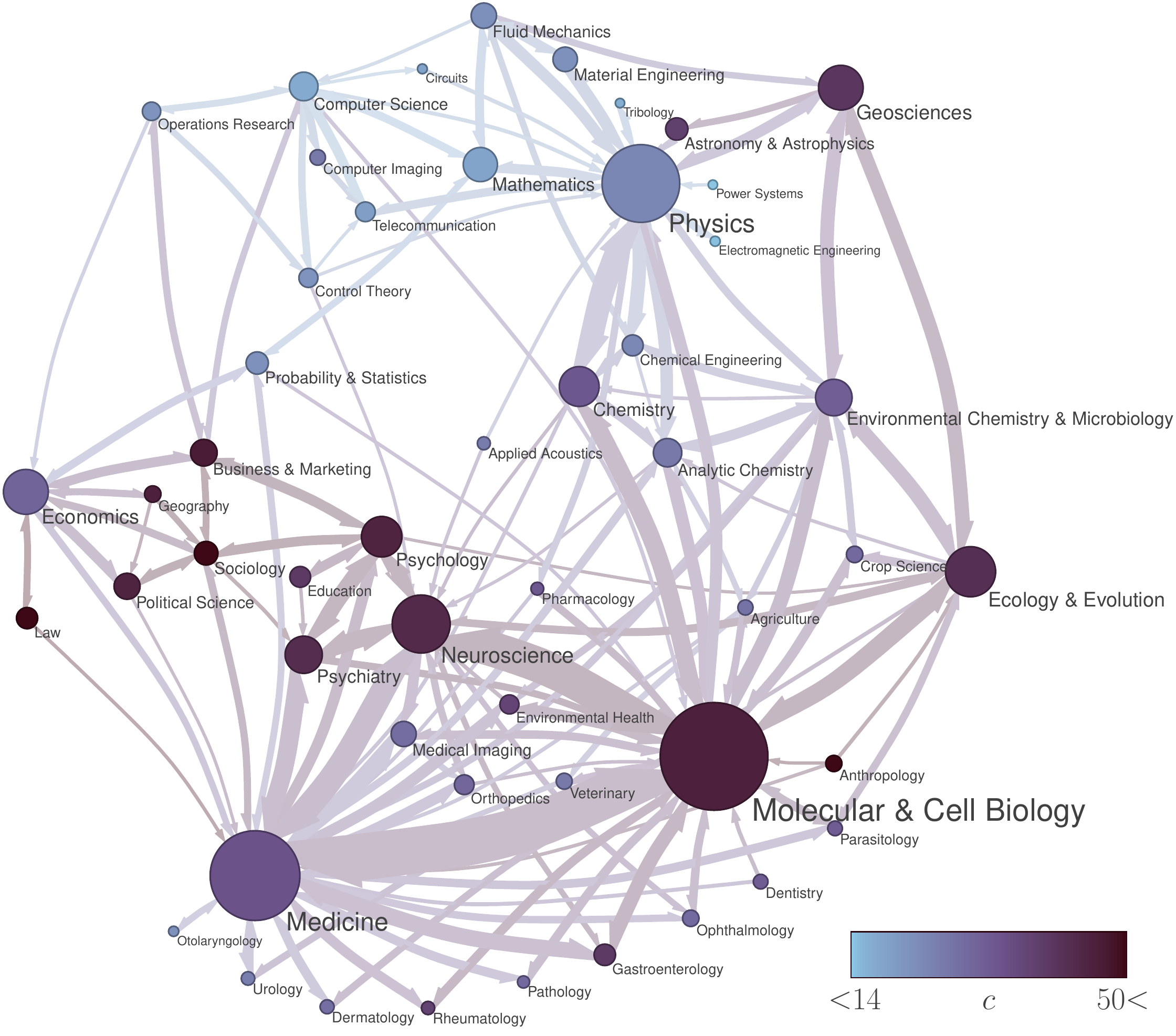}\hskip .25in
\includegraphics[width=3.3 in]{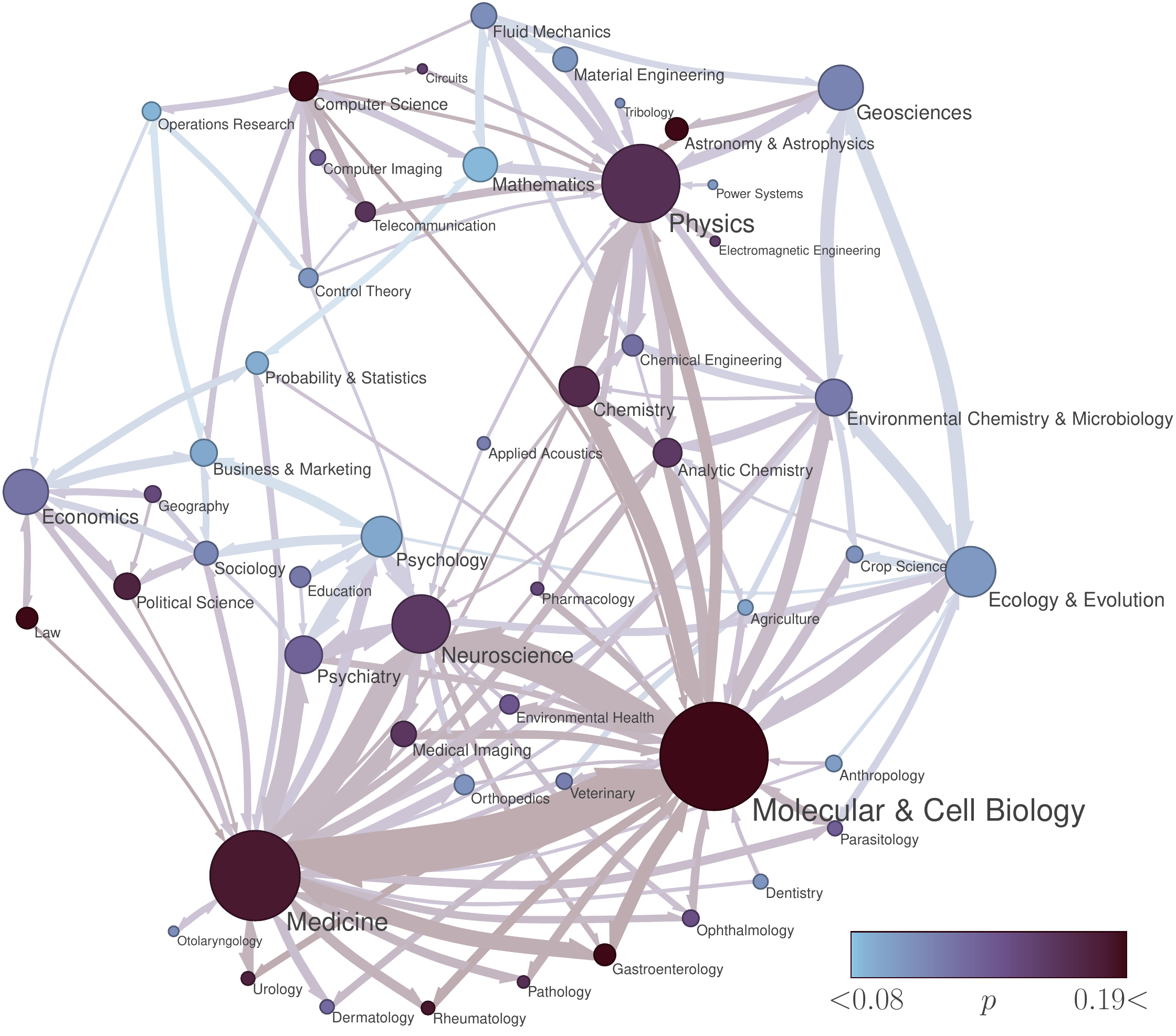}
\vskip .25in
\includegraphics[width=3.3 in]{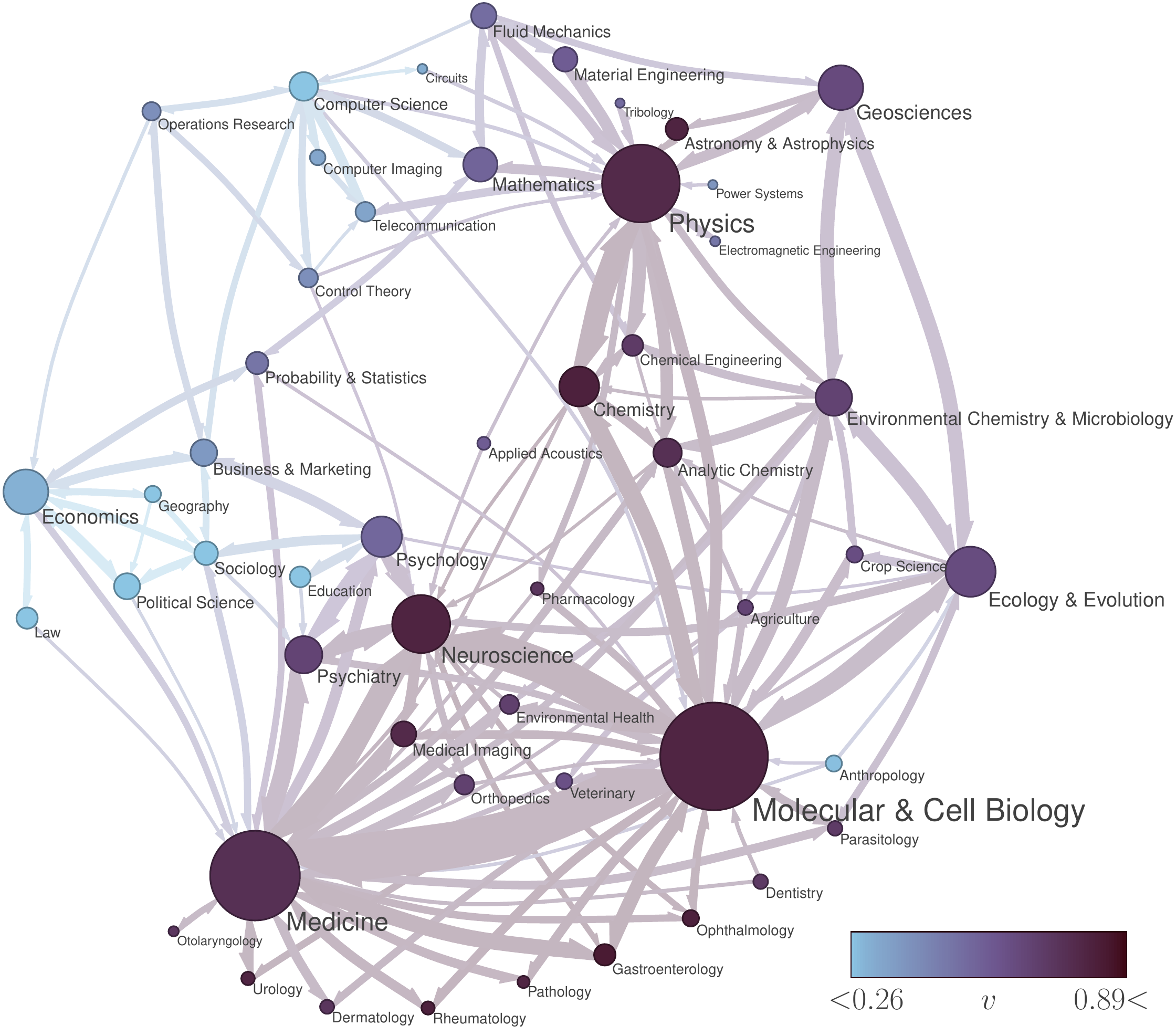}\hskip .25in
\includegraphics[width=3.3 in]{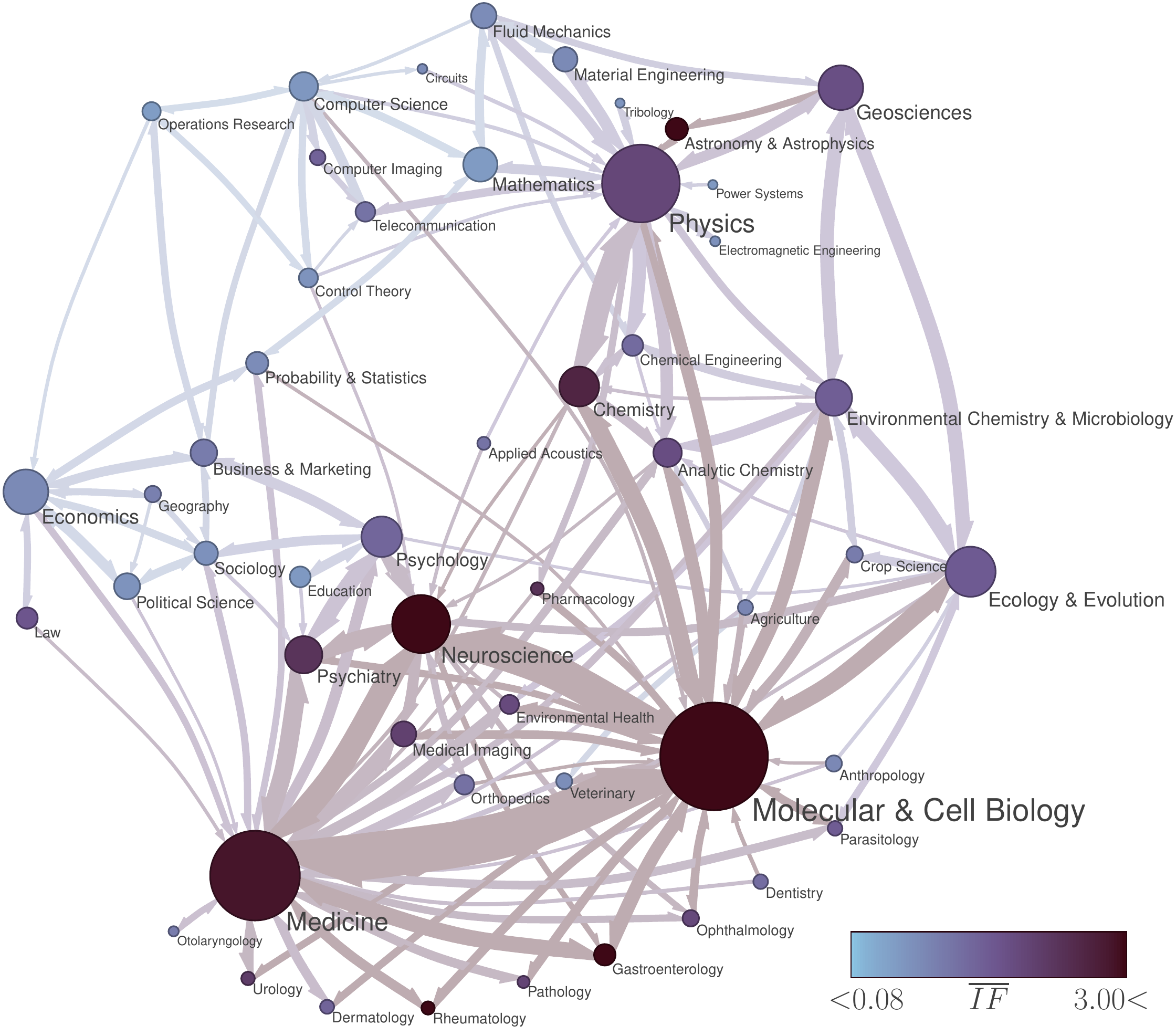}
\caption{{\bf Differences in citation pattern across fields.} Fields are categorized and mapped as in Rosvall and Bergstrom (2008). Panel a: average number of items cited per paper. Panel b: Fraction of citations to papers published in the two previous calendar years. Panel c: Fraction of citations to papers published in JCR-listed journals. Panel d: Weighted impact factor, $\overline{\text{IF}}$.
}
\label{maps}
\end{figure*}

Figure \ref{maps} summarizes the differences in weighted average impact factor across fields (panel d) and the factors responsible for these differences (panels a--c).

\subsection{Inflation differences across fields}

As we have shown in previous sections, weighted impact factor is increasing every year and is different for each field. Naturally, the next several questions to be asked are \textit{Is inflation ubiquitous across fields? Do some fields inflate more than others? Which fields inflate the most?}. Differences in inflation rates between fields will be important when evaluating citation data within a specific field over time. Knowing that, for instance, psychiatry is inflating twice as fast as neuroscience, would help one compare journals across these fields over time.

The results of the analysis are reported in Table \ref{fieldcompare}. Fields vary substantially in their rates of impact factor inflation. Further analysis shows that inflation rate is not correlated to size of field ($r^2=0.001$), nor weighted impact factor scores of that field ($r^2=0.018$).

\begin{table*}[htdp]
\centering
\begin{tabular}{l llllll}
& &&&& &Growth\\
Field (Size)& $\overline{\text{IF}}$ & $\alpha$ & $c$& $p$& $v$& Rate\\
\hline
Molecular and Cell Biology (511) \hspace{.2in} & 4.763 \hspace{.2in} & 0.515 \hspace{.2in} & 45.810 \hspace{.2in} & 0.205 \hspace{.2in} & 0.803 \hspace{.2in} & 0.006\\
Astronomy and Astrophysics (25) & 4.295 & 0.530 & 38.249 & 0.215 & 0.813 & 0.074\\
Gastroenterology (40) & 3.475 & 0.494 & 39.669 & 0.193 & 0.849 & 0.030\\
Rheumatology (20) & 3.348 & 0.519 & 37.818 & 0.184 & 0.826 & 0.079\\
Neuroscience (224) & 3.252 & 0.515 & 43.768 & 0.159 & 0.810 & 0.017\\
Medicine (766) & 2.896 & 0.515 & 33.920 & 0.183 & 0.760 & 0.036\\
Chemistry (145) & 2.610 & 0.539 & 33.103 & 0.170 & 0.821 & 0.026\\
Pharmacology (28) & 2.331 & 0.575 & 32.947 & 0.149 & 0.737 & 0.098\\
Psychiatry (178) & 2.294 & 0.522 & 43.025 & 0.131 & 0.670 & 0.039\\
Urology (23) & 2.132 & 0.513 & 25.501 & 0.176 & 0.806 & 0.032\\
Medical Imaging (84) & 2.043 & 0.502 & 28.727 & 0.161 & 0.784 & 0.034\\
Pathology (28) & 1.991 & 0.516 & 29.523 & 0.166 & 0.803 & 0.020\\
Physics (503) & 1.912 & 0.543 & 23.963 & 0.167 & 0.783 & 0.018\\
Ophthalmology (36) & 1.905 & 0.536 & 29.105 & 0.144 & 0.823 & 0.029\\
Environmental Health (73) & 1.871 & 0.533 & 37.234 & 0.140 & 0.691 & 0.048\\
Analytic Chemistry (129) & 1.789 & 0.538 & 26.702 & 0.158 & 0.762 & 0.022\\
Geosciences (224) & 1.768 & 0.526 & 40.529 & 0.113 & 0.647 & 0.021\\
Law (71) & 1.657 & 0.485 & 76.826 & 0.199 & 0.231 & 0.010\\
Ecology and Evolution (349) & 1.555 & 0.523 & 42.172 & 0.100 & 0.640 & 0.051\\
Parasitology (38) & 1.527 & 0.505 & 32.076 & 0.134 & 0.711 & 0.036\\
Environmental Chemistry & 1.505 & 0.518 & 31.648 & 0.117 & 0.679 & 0.039\\
and Microbiology (181) \\
Computer Imaging (31) & 1.446 & 0.514 & 26.470 & 0.133 & 0.332 & 0.067\\
Dermatology (38) & 1.427 & 0.480 & 28.442 & 0.128 & 0.734 & 0.050\\
Psychology (210) & 1.387 & 0.513 & 45.139 & 0.091 & 0.538 & 0.033\\
Chemical Engineering (75) & 1.290 & 0.587 & 23.660 & 0.124 & 0.711 & 0.041\\
Dentistry (43) & 1.284 & 0.529 & 32.046 & 0.102 & 0.717 & 0.029\\
Orthopedics (72) & 1.226 & 0.531 & 30.033 & 0.103 & 0.683 & 0.066\\
Telecommunication (37) & 1.192 & 0.550 & 19.518 & 0.163 & 0.334 & 0.054\\
Applied Acoustics (36) & 1.171 & 0.526 & 25.942 & 0.115 & 0.575 & 0.031\\
Crop Science (61) & 1.040 & 0.523 & 29.467 & 0.104 & 0.631 & 0.025\\
Business and Marketing (101) & 1.035 & 0.538 & 46.865 & 0.091 & 0.376 & 0.032\\
Geography (56) & 0.986 & 0.526 & 46.055 & 0.148 & 0.254 & 0.029\\
Information Science (23) & 0.918 & 0.539 & 28.402 & 0.220 & 0.217 & 0.078\\
Agriculture (56) & 0.882 & 0.530 & 27.503 & 0.093 & 0.670 & 0.024\\
Anthropology (62) & 0.872 & 0.496 & 52.104 & 0.098 & 0.275 & 0.020\\
Material Engineering (107) & 0.826 & 0.537 & 22.038 & 0.100 & 0.578 & 0.063\\
Economics (159) & 0.823 & 0.511 & 30.423 & 0.121 & 0.299 & 0.021\\
Fluid Mechanics (107) & 0.804 & 0.520 & 22.096 & 0.107 & 0.516 & 0.041\\
Probability And Statistics (57) & 0.796 & 0.528 & 21.974 & 0.089 & 0.496 & 0.023\\
Veterinary (77) & 0.767 & 0.480 & 26.512 & 0.115 & 0.620 & 0.041\\
Sociology (96) & 0.715 & 0.510 & 50.840 & 0.110 & 0.189 & 0.001\\
Media and Communication (24) & 0.690 & 0.479 & 46.932 & 0.133 & 0.190 & 0.024\\
Control Theory (64) & 0.681 & 0.474 & 21.394 & 0.102 & 0.407 & 0.061\\
Political Science (99) & 0.680 & 0.500 & 45.014 & 0.176 & 0.131 & 0.012\\
Computer Science (124) & 0.631 & 0.717 & 17.215 & 0.193 & 0.266 & 0.034\\
Education (86) & 0.590 & 0.509 & 39.890 & 0.119 & 0.213 & 0.015\\
Mathematics (149) & 0.556 & 0.512 & 18.477 & 0.085 & 0.552 & 0.033\\
Operations Research (62) & 0.542 & 0.521 & 21.714 & 0.086 & 0.408 & 0.043\\
History and Philosophy & 0.456 & 0.507 & 51.316 & 0.068 & 0.159 & -0.003\\
Of Science (32) \\
History (23) & 0.416 & 0.466 & 81.775 & 0.101 & 0.059 & -0.028
\end{tabular}
\caption{Table showing $\alpha, \ c,\ p,\ v$ and exponential growth rates for individual fields. All except growth rate were calculated using 2004 data.}
\label{fieldcompare}
\end{table*}

\section*{Summary}

Impact factors vary across fields and over time. By decomposing average impact factors into four contributing components --- field growth, average number of cited items per paper, fraction of citations to papers published within two years, and fraction of citations to JCR-listed items --- we are able to determine the sources of this variation. We find that an increasing number of citations in the reference lists of published papers is the greatest contributor to impact factor inflation over time. Differences in the fraction of citations to JCR-indexed literature is the greatest contributor to differences across fields, though cross-field differences in impact factor are also influenced by differences in the number of citations per paper and differences in the fraction of references that were published within two years.  By contrast, the growth rate of the scientific literature and cross-field differences in net size and growth rate have very little influence on impact factor inflation or on cross-field differences in impact factor. 

\section*{Competing interests}
The authors are the developers of Eigenfactor (\texttt{http://www.eigenfactor.org}), a method for ranking journal influence using citation network data. 

\section*{Acknowledgments}
The authors would like to thank Martin Rosvall for generating the maps used in Figure \ref{maps}, and also Alan Wilson for drawing out attention to impact factor inflation. This work was supported in part by a Mary Gates Research Scholarship and a Howard Hughes Medical Institute Integrative Research Internship to B.M.A.

\appendix

\section{Deriving $\alpha_t$, $c_t$, $p_t$ and $v_t$ from the JCR data}
\label{Appendix:methods}

All citation data sets come from the JCR data sets for the years 1994 through 2005. The JCR does not list article counts for year $t$ in data set for year $t$; the year $t+1$ and year $t+2$ data sets typically do not agree exactly on the number of articles that were published in year $t$.  Therefore, in order to compute the the year $t$ article count, A$_t$, we average the article count listed for year $t$ in the $t+1$ data set and year $t$ in the $t+2$ data set.  We then calculate $\alpha_t = \text{A}_t / (\text{A}_{t-1} + \text{A}_{t-2})$ using the total article counts for years $\text{A}_{t-1}$ and $\text{A}_{t-2}$ as given in the data set for year $t$.

We calculate $c_t$ by dividing the total outgoing citations for all journals in year $t$ by the total articles for year $t$:
 \begin{equation*}
 c_t = \frac{\text{total out-citations in year }t}{\text{A}_t}.
 \end{equation*}

 We calculate $p_t$ by dividing the total outgoing citations for all journals to material published in the previous two years ($t-1$ and $t-2$) by the total outgoing citations for all journals in year $t$:
 \begin{equation*}
 p_t = \frac{\text{2-year total out-citations from year }t}{\text{total out-citations in year }t}.
 \end{equation*}

\begin{figure}[htdb]
\centering
\includegraphics[width=2.8 in]{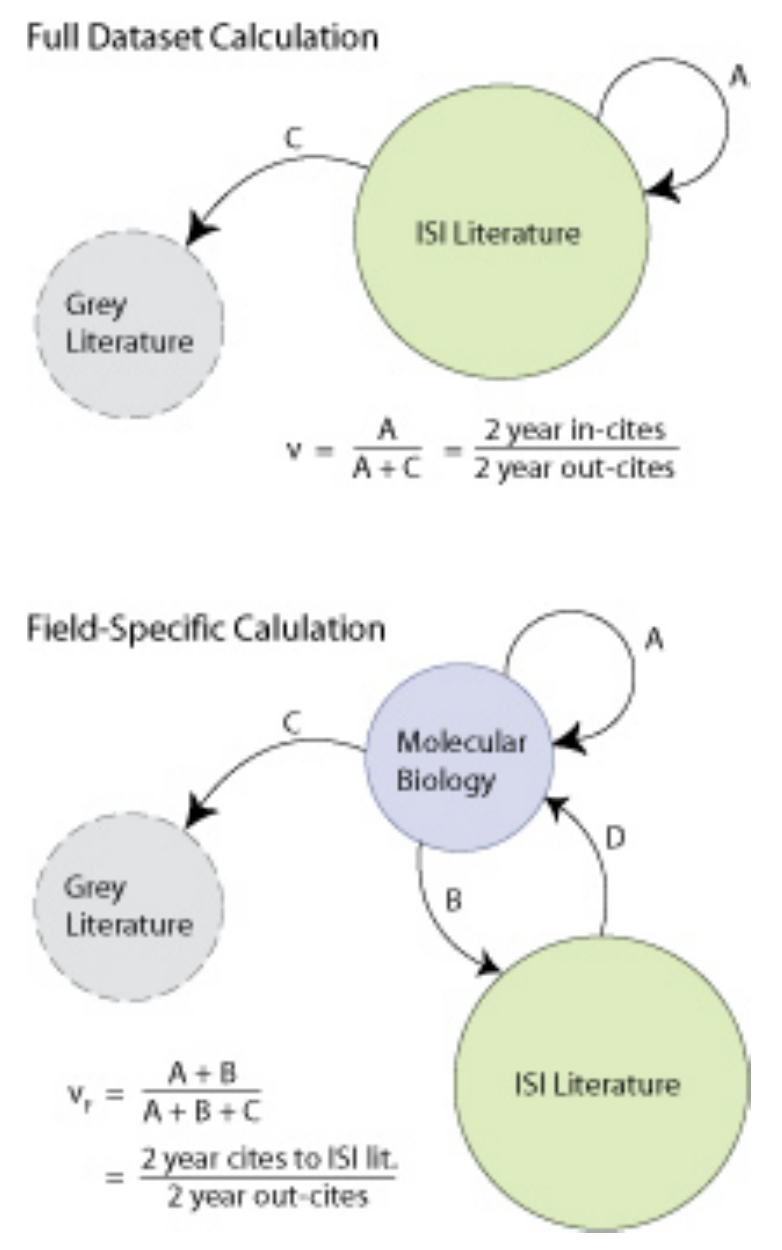}
\caption{{\bf Calculating $v_t$.} Top panel gives the schematic for calculating $v_t$ for the entire dataset, and the bottom panel gives the schematic for specific fields.}
\label{vdiagram}
\end{figure}

The calculation of $v_t$ is slightly more complicated than the other calculations; Figure \ref{vdiagram} provides a schematic representation. To calculate the percentage of citations into the JCR for the entire dataset we divide the total {\em incoming} citations for the previous two years (figure \ref{vdiagram}, top panel, arrow A) by the total outgoing citations over that period (arrows A + C):

\begin{equation*}
v_{(t, \text{Entire Dataset})} = \frac{\text{2-year total in-citations from year }t}{\text{2-year total out-citations from year }t}. \end{equation*}
 
 \noindent This is done because the incoming citations for the entire dataset are the outgoing citations from the JCR to itself. However, this is not true for the specific field calculations. To calculate $v_{t, F}$ for any field $F$, we divide the 2-year outgoing citations from that field to itself (figure \ref{vdiagram}, bottom panel, arrow A) plus the 2-year outgoing citations from that field to the rest of the JCR (arrow B) by the total 2-year outgoing citations from that field (arrows A + B + C):

\begin{equation*}
v_{(t,F)} = \frac{
\left(
\begin{tabular}{ l }
\text{2-year out-citations from $F$ to $F$ +} \\
\text{2-year out-citations into rest of JCR} \\
\end{tabular}\right)}{\text{2-year total out-citations from year }t}. \end{equation*}

\end{document}